\documentclass[prl,twocolumn,showpacs,letterpaper,showpacs,superscriptaddress]{revtex4}
\usepackage{graphicx,amsmath,amssymb,amsfonts,latexsym,color,dcolumn,bm,epsfig,subfigure}

\begin{document}

\title{Contribution of drifting carriers to the Casimir-Lifshitz and Casimir-Polder interactions with semiconductor materials}

\author{Diego A. R. Dalvit}
\affiliation{Theoretical Division, Los Alamos National Laboratory, Los Alamos, NM 87545, USA}

\author{Steve K. Lamoreaux}
\affiliation{Yale University, Department of Physics,
P.O. Box 208120, New Haven, CT 06520-8120, USA}

\date{\today}

\begin{abstract}

We develop a theory for Casimir-Lifshitz and Casimir-Polder interactions with semiconductor or insulator surfaces that takes into account charge drift in the bulk material through use of the classical Boltzmann equation.
We derive frequency-dependent dispersion relations that give the usual Lifshitz results for dielectrics as a limiting case and, in the quasi-static limit,
coincide with those recently computed to
account for Debye screening in the thermal Lifshitz force with conducting surfaces with small density of carriers.
\end{abstract}


\maketitle

{\it Introduction.-} Propagating waves inside semiconductors interact with drifting carriers in the bulk material,
and this is the basis of phenomena such as solid-state traveling-wave amplification
\cite{Sumi,Thiennot}. Typically, an ultrasonic wave or a microwave incident on a semiconductor is amplified
when the mean drift velocity of carriers exceeds the phase velocity of the propagating wave.
The theoretical description of this phenomenon involves
Maxwell's equations for the electromagnetic field coupled to the classical Boltzmann transport equation
to describe the motion of charged carriers in the bulk semiconductor.

In principle, the same type of coupling between propagating waves and drifting carriers
is also present for quantum vacuum fluctuations of the electromagnetic field in the presence
of semiconductor boundaries. Hence, one should expect that the complete description of the
Casimir-Lifshitz force between bulk materials and the atom-surface Casimir-Polder force
\cite{Lifshitz} should take into account the possibility of carrier drift when one of the surfaces involved is a semiconductor or a conductor with small density of carriers. In this limit, the classical Boltzmann equation can be used to determine the dynamic equilibrium between a time- and spatially-varying electric field and changes in the charge density within the material.

The effect of material properties on quantum vacuum forces is encapsulated in
the Lifshitz theory through the frequency-dependent reflection amplitudes $r^p_{{\bf k},j}(i\xi_n)$ of the $j$-th material boundary. Here $p$ denotes the polarization of incoming waves (transverse electric TE or
transverse magnetic TM), ${\bf k}$ is their transverse momentum, and
the reflection amplitudes are evaluated at imaginary frequencies $\omega=i \xi_n$, where
$\xi_n= 2 \pi n k_{\rm B} T / \hbar$ are the Matsubara frequencies.
The Casimir-Lifshitz pressure between two plane semi-spaces separated by a vacuum gap $d$
is \cite{Lifshitz}
\begin{equation}
P(d)= 2 k_{\rm B} T
{\sum_{n=0}^{\infty}}'\int \frac{d^2 {\bf k}}{(2 \pi)^2}
\, K_3 \, \sum_p
\frac{r^p_1 r^p_2 e^{-2K_3 d} }{ 1- r^p_1 r^p_2 e^{-2K_3 d}},
\label{CL}
\end{equation}
where $K_3 = \sqrt{k^2 + \xi^2/c^2}$ and the prime in the sum over $n$ means that a factor 1/2 is to be included for the $n=0$ term.  Assuming that one of the
media is dilute, one can derive from Eq.(\ref{CL}) the Casimir-Polder force on an atom above a planar surface \cite{Lifshitz}.
It has been shown that the form of the plate(s) electrical permittivity used
to compute the reflection amplitudes via Fresnel relations
vastly alter the magnitude and form of the force in these calculations
\cite{sernelius}. The effect of carrier drift in the case of dynamic fields
has not yet been studied in relation to Casimir-like forces, and as we show in this Letter, alters the form of the field mode equations.

Recently Pitaevskii \cite{Pitaevskii08} has proposed a theory for the thermal
Lifshitz force between an atom and a conductor with a small density of carriers that
takes into account the penetration of the static component of the fluctuating EM
field into the conductor. This approach is quasi-static, appropriate for the
large distance regime of the thermal Lifshitz atom-surface interaction, and is essentially
based on the Debye-H\"uckel charge screening \cite{Landau}. In this static limit, the
reflections coefficients are $r^{\rm TE}_{{\bf k}}(0)=0$ and
$r^{\rm TM}_{{\bf k}}(0)=(\overline{\epsilon}_0 q-k)/(\overline{\epsilon}_0 q +k)$.
Here $\overline{\epsilon}_0$ is the static ``bare" dielectric constant of the medium (which
does not take into account the contribution from current carriers),
$q =\sqrt{k^2 + \kappa^2}$, and
$\kappa^2= 4 \pi e^2 n_0 / \overline{\epsilon}_0 k_{\rm B} T$, where $-e$ is the electron charge and
$n_0$ is the (uniform) carrier density \cite{Pitaevskii08}. Note that
$\kappa=1/R_{\rm D}$ is the inverse of the Debye radius $R_{\rm D}$. For good metals the Debye
radius is very small (on the order of inter-atomic distances), while for semiconductors it is much larger (on the order of microns or more).
This quasi-static calculation  for the thermal Lifshitz force
interpolates between the ideal dielectric limit ($d \ll R_{\rm D}$) and the
good conductor limit ($d \gg R_{\rm D}$).
We further point out that the Debye-H\"uckel charge screening effect can produce
a large correction to an electrostatic calibration because a static field can
penetrate a finite distance into the plates, leading to an error in the
determination of the plate separation \cite{Lamoreaux08}. On the other hand, it can be  expected that screening should affect dynamic fields as well; however the phenomenological dispersion relation suggested in \cite{Lamoreaux08} for dynamic fields
is shown here to be incorrect.  In the following we will
extend Pitaevskii's calculation beyond the quasi-static regime, and compute
the frequency-dependent TE and TM reflection coefficients $r^p_{{\bf k}}(i\xi)$ for semiconductor media taking into account carrier drift.



{\it Field equations .-}
For an intrinsic semiconductor, the densities of carriers
and holes are comparable, but the dynamics are different. Here
we follow the approach in \cite{Sumi,Thiennot}, and treat
the carriers and holes as dynamically equivalent, which roughly doubles the charge density. This treatment is very accurate
in the quasi-static limit.
Assuming that
there is no external applied field on the semiconductor, and
all fields have a time
dependency of the form $e^{-i \omega t}$, Maxwell's equations take the form
$\nabla \times {\bf E} = i \mu_0 \omega {\bf H}$,
$\nabla \times {\bf H} = -i \overline{\epsilon}(\omega) \omega {\bf E} + {\bf J}$,
and $\nabla \cdot {\bf E} = - e n/ \overline{\epsilon}(\omega)$. Here
$\overline{\epsilon}(\omega)$ is the  frequency-dependent ``bare" permittivity of the semiconductor, that does not take into account the contribution from current carriers,
$n$ is the intrinsic carrier density, and
$\mu_0$ is the permeability of vacuum.
The carrier current is ${\bf J}=-e n {\bf v}$, where
${\bf v}$ is the mean velocity of carriers.
The fact that carriers can drift in the semiconductor is modelled by Boltzmann transport equation \cite{Sumi,Thiennot}
\begin{equation}
\left(
\frac{\partial}{\partial t} + {\bf v} \cdot \nabla \right) {\bf v} =
-\frac{e}{m} {\bf E} - \frac{v_T^2}{n} \nabla n - \frac{{\bf v}}{\tau} ,
\label{Boltzmann}
\end{equation}
where $m$ is the effective mass of the charge carriers, $v_T=\sqrt{k_{\rm B} T/m}$ is their mean thermal velocity, and $\tau$ is the carrier relaxation time.
Linearizing Eq.(\ref{Boltzmann}) with respect to the ac fields which have a
factor $e^{-i \omega t}$, one can solve it for $\nabla n$ and then use the result in Maxwell's equation to derive
the fundamental equation for the electric field inside the semiconductor
\cite{Sumi,Thiennot}
\begin{equation}
\left[ \nabla^2 + \mu_0 \overline{\epsilon}(\omega) \omega^2 \left(1+ i \frac{\tilde{\omega}_c}{\omega}\right) \right] {\bf E} = [ 1 + i \mu_0 \overline{\epsilon}(\omega) \omega \tilde{D} ] \nabla
\cdot (\nabla \cdot {\bf E}) .
\label{main}
\end{equation}
Here $\tilde{\omega}_c = \omega_c / (1-i \omega \tau)$ and
$\tilde{D} = D/(1-i \omega \tau)$, where
$\omega_c  = 4 \pi e n_0 \mu/\overline{\epsilon}(\omega)$,
$\mu=e \tau/m$ is the mobility of carriers,
and $D=v_T^2 \tau$ is the diffusion constant.
Note that the frequency-dependent ratio
$\omega_c/D= 4 \pi e^2 n_0/ \overline{\epsilon}(\omega) k_{\rm B} T$ coincides with
$\kappa^2=1/R_{\rm D}^2$ in the quasi-static limit.



{\it TM and TE reflection amplitudes.-}
Let us assume that the semiconductor occupies the semi-space region $z<0$ and the region $z>0$ is vacuum. Eq.(\ref{main}) allows TM and TE solutions. For TM modes $e_y=0$, so that (the phase factors $e^{-i \omega t}$ will be omitted from now on)
${\bf E} = [ e_x(z) {\hat{\bf x}} + e_z(z) {\hat{\bf z}} ]  e^{i k x}$.
Substituting this into Eq.(\ref{main}) one gets two coupled equations, which can be combined into two
uncoupled fourth-order differential equations for $e_x$ and $e_z$, namely
$( \partial^2_z - \eta^2_T) (\partial^2_z - \eta^2_L) e_x =0$ and
$( \partial^2_z - \eta^2_T) (\partial^2_z - \eta^2_L) e_z = 0$,
where \cite{Sumi}
\begin{eqnarray}
\eta^2_T & = & k^2 - \mu_0 \overline{\epsilon}(\omega) \omega^2 \left( 1+ i \frac{\tilde{\omega}_c}{\omega} \right) , \\
\eta^2_L & = & k^2 - i \frac{\omega}{\tilde{D}}
\left( 1+ i \frac{\tilde{\omega}_c}{\omega} \right) .
\end{eqnarray}
The solutions that vanish for $z \rightarrow - \infty$ are
$e_x(z) = A_T e^{\eta_T z} + A_L e^{\eta_L z}$ and
$e_z(z) = A'_T e^{\eta_T z} + A'_L e^{\eta_L z}$,
where we assume ${\rm Re}\; \eta_T$ and ${\rm Re}\;  \eta_L$ to be positive. The amplitudes $A_L$ and $A_T$ are arbitrary so far, and
$A'_L = -i \eta_L A_L / k$ and $A'_T=-i k A_T / \eta_T$. The magnetic field inside the semiconductor is
${\bf H} = i {\hat{\bf y}} A_T e^{\eta_T z} e^{i k x} (k^2- \eta_T^2)/ \mu_0 \omega \eta_T$.

The boundary conditions on the $z=0$ interface are
${\bf H}_{\|}$, ${\bf E}_{\|}$, ${\bf D}_{\bot}$ and
${\bf B}_{\bot}$ continuous. The latter one is automatically satisfied for TM modes, while the other ones imply
$E_x$ and $H_y$ continuous, and $\overline{\epsilon}(\omega) E_z$ continuous. Imposing these boundary conditions, and using the expressions for the fields inside the semiconductor derived above, we obtain the reflection amplitude for fields
impinging from the vacuum side,
$r= (1-\alpha)/(1+\alpha)$,  with
$\alpha = \frac{k^2}{i \eta_L k_z} \left[
\frac{1}{\overline{\epsilon}(\omega)} - \frac{\omega^2/c^2}{k^2-\eta^2_T} +
\frac{\eta_L \eta_T \omega^2/c^2}{k^2 (k^2-\eta^2_T)}
\right]$. Expressed along imaginary frequencies
$\omega=i \xi$, the TM reflection amplitude is
\begin{equation}
r^{\rm TM}_{{\bf k}}(i\xi) =
\frac{\overline{\epsilon} (i \xi) \sqrt{k^2 + \xi^2/c^2} - \chi}{\overline{\epsilon}(i \xi) \sqrt{k^2 + \xi^2/c^2} + \chi},
\label{refTM}
\end{equation}
where
$\chi = \frac{1}{\eta_L} \left[ k^2 + \overline{\epsilon}(i \xi) \frac{\xi^2}{c^2}
\frac{\eta_L \eta_T - k^2}{\eta_T^2 - k^2} \right]$.

For TE modes $e_z=0$, so that
${\bf E} = [ e_x(z) {\hat{\bf x}} + e_y(z) {\hat{\bf y}} ]  e^{i k x}$.
Plugging this into Eq.(\ref{main}) one gets two equations:
$[\partial^2_z + \mu_0 \overline{\epsilon}(\omega) \omega^2 (1+i \tilde{\omega}_c/\omega + i \tilde{D} k^2/\omega)] e_x = 0$, and
$[\partial^2_z -k^2 + \mu_0
\overline{\epsilon}(\omega) \omega^2 (1+i \tilde{\omega}_c/\omega)] e_y = i k [1+i \mu_0 \overline{\epsilon}(\omega) \omega \tilde{D}] \partial_z e_x$. The solutions are $e_x(z) =  A e^{\beta z}$ and $e_y(z) = B e^{\eta_T z} + C e^{\beta z}$. Here $A$ and  $B$ are constants, $\beta^2= -i \mu_0 \overline{\epsilon}(\omega) \omega \tilde{D} \eta_L^2$, and
$C=i k A \beta (1+ i \mu_0 \overline{\epsilon}(\omega) \omega \tilde{D}) / (\beta^2 - \eta^2_T)$ (we assume ${\rm Re} \beta > 0$).
The magnetic field inside the semiconductor is
${\bf H} =(1/i \mu_0 \omega)
[- (B \eta_T e^{\eta_T z} + C \beta e^{\beta z}) {\hat{\bf x}}
+ A \beta e^{\beta z} {\hat{\bf y}}
+ i k (B \eta_T e^{\eta_T z} + C \beta e^{\beta z}) {\hat{\bf z}} ]
e^{i k x}$.

Imposing the boundary conditions on the $z=0$ interface,
and upon performing the rotation $\omega\rightarrow i \xi$,
we get the TE reflection amplitude for fields impinging from
the vacuum side
\begin{equation}
r^{\rm TE}_{{\bf k}}(i\xi) =
\frac{\sqrt{k^2 + \xi^2/c^2} - \eta_T}{\sqrt{k^2 + \xi^2/c^2} + \eta_T}.
\label{refTE}
\end{equation}
Note that $\eta_T^2=k^2 + [\overline{\epsilon}(i \xi) + 4 \pi \sigma(i \xi)/\xi] \xi^2/c^2$ (where
$\sigma(i\xi)=\sigma_0/(1+\xi \tau)$ and $\sigma_0=e^2 n_0 \tau/m$ are the ac and dc Drude conductivities, respectively), so
Eq.(\ref{refTE}) gives the usual Fresnel TE reflection coefficient with account of ac Drude conductivity.
On the other hand, Eq.(\ref{refTM}) gives a modified Fresnel TM reflection coefficient due to the
presence of Debye-H\"uckel screening and charge drift.



{\it Limiting cases.-} Let us study the behavior of the frequency-dependent reflection amplitudes we have derived above for some interesting limiting cases.
(a) {\it Quasi-static limit:} When $\xi \rightarrow 0$, we have
$\eta_T^2 \approx k^2+ \overline{\epsilon}_0  \xi \omega_c/c^2$ and $\eta_L^2 \approx k^2 + \kappa^2 = q^2$ (recall that
$\omega_c/D= \kappa^2$ in the quasi-static limit). Here $\overline{\epsilon}_0 \equiv \overline{\epsilon}(0)$. Therefore,
$\chi \approx k^2/q$, and from (\ref{refTM}) and (\ref{refTE}) we obtain
that the zero-frequency limit of the reflection amplitudes is
$r^{\rm TE}_{{\bf k}} (0) = 0$ and $r^{\rm TM}_{{\bf k}}(0)= (\overline{\epsilon}_0 q - k)/(\overline{\epsilon}_0 q + k)$,
which coincides with the prediction of \cite{Pitaevskii08} for the reflection
coefficients in the quasi-static limit. In consequence, we recover the correct
thermal Lifshitz force between an atom and a surface with small density of carriers, and the associated crossover between good conductors
and ideal dielectrics.
%
%
(b) {\it Ideal dielectric limit:} In this case
the free charge density is small ($n_0 \ll 1$), and the discrete charges are quasi-bound, making their effective thermal velocity very small.
Therefore,  $\tilde{\omega}_c / \tilde{D}\approx 1/\lambda_{\rm D}^2$ is small  (as in the quasi-static limit for ideal dielectrics), where
$\tilde{\omega}_c \approx 4 \pi e^2 n_0/m \xi \overline{\epsilon}(i\xi)$ and
$\tilde{D}$ are both small.
Consequently $\eta_T^2 \approx k^2 + \overline{\epsilon}(i\xi) \xi^2/c^2$,
$\eta_L^2 \approx \xi^2/v_T^2 \rightarrow \infty$, and
$\chi \approx \eta_T$. We recover from (\ref{refTM}) and (\ref{refTE})
the usual expressions for the reflection coefficients for ideal dielectrics.



\begin{figure}[t]
\scalebox{1.2}{\includegraphics{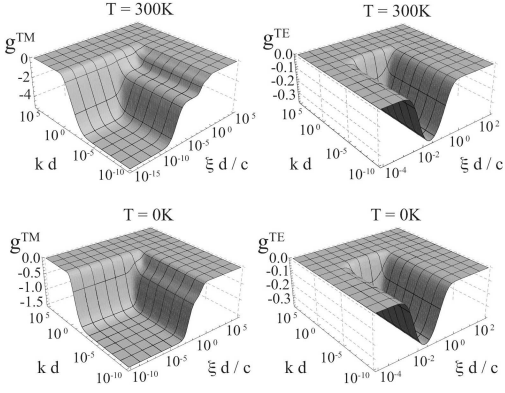}}
\caption{Behavior of the functions $g^p_{\bf k}(i\xi)$ used to compute the Casimir-Lifshitz free energy and entropy for semiconductor materials with account of drifting carriers.
The reflections coefficients are given by (\ref{refTM}) and (\ref{refTE}),
parameters are for intrinsic Ge (see text), and the
distance is set to $d=1\mu$m. The variation with temperature
(in the range $T=0-300$K) of the TE function is not perceptible on the scale of the figure. The
corresponding functions without account of Debye screening and carrier drift correspond to the $T=0$K plots in this figure.}
\end{figure}


{\it Free energy and entropy.-}
The Casimir-Lifshitz free energy for two parallel planar media is
\begin{eqnarray}
&& E = \frac{A \hbar}{2 \pi} \sum_{p} \sum_{n=0}^{\infty'}
\int \frac{d^2{\bf k}}{(2 \pi)^2} \;
\theta \; g^{p}_{{\bf k}}(i n \theta,\theta), \\
&&g^{p}_{{\bf k}}(\omega,\theta)=\ln[1- r_{{\bf k},1}^p(\omega,\theta) r_{{\bf k},2}^p(\omega,\theta) e^{-2 d \sqrt{k^2-\omega^2/c^2}}] \nonumber,
\end{eqnarray}
where  $\theta=2 \pi k_{\rm B} T/\hbar$ and $A$ is the area of the plates.  Note that we have allowed  for an explicit dependence of the reflection coefficients on temperature. In Fig.1 we plot the behavior of
$g^p_{{\bf k}}$ as a function of the imaginary frequency $\omega=i\xi$ and transverse momentum
$k$ for TM and TE polarizations (the corresponding reflection amplitudes are obtained from
Eqs.(\ref{refTM}) and (\ref{refTE})). As an example, we consider the case of two identical media made of intrinsic germanium. The permittivity of Ge
is known to have a weak dependence on temperature, and becomes constant as $T$ goes to zero \cite{index}.
It can be approximately fitted with a Sellmeier-type expression $\overline{\epsilon}(i\xi)=\overline{\epsilon}_{\infty} + \omega_0^2 (\overline{\epsilon}_0-\overline{\epsilon}_{\infty}) / (\xi^2+\omega_0^2)$, with
$\overline{\epsilon}_0 \approx 16.2$, $\overline{\epsilon}_{\infty} \approx 1.1$,
and $\omega_0 \approx 5.0 \times 10^{15}$ rad/sec. The intrinsic carrier density varies with temperature as $n_0(T)=\sqrt{n_c n_v} e^{-E_g/2 k_{\rm B} T}$, where $E_g$ is the energy gap,
and $n_c$ ($n_v$) is the effective density of states in the conduction (valence) band
\cite{Gedata}. The relaxation time $\tau$ has an exponential dependency on temperature,
and at low temperatures goes linearly in $T$ to a non-zero constant \cite{Gedata}.
Given typical parameters of intrinsic semiconductors, $\tilde{\omega}_c$ and $\tilde{D}/\xi$ are both very small in the relevant range of frequencies for the Lifshitz formula, and then only the $n=0$ TM mode is modified significantly. The effect of drifting carriers can therefore, to very high accuracy, be fully modeled by the Debye-H\"uckel screening length.

Our theory for Casimir forces taking into account the possibility of carrier drift
in intrinsic semiconductor media is compatible with Nernst theorem of thermodynamics,
that states that the entropy $S=- dE/dT$
should vanish at zero temperature for a system with a nondegenerate ground state.
Whether the systems we are considering here have nondegenerate ground states remains open; however, satisfaction of the Nernst theorem provides weak evidence for the possible viability of a theoretical model.  Following, for example,  the technique in \cite{Intravaia},
and using the fact that the intrinsic carrier density vanishes exponentially as $T \rightarrow 0$
(which in turn implies that the derivatives of $g^p_{\bf k}(i \xi, \theta)$ with respect to $\xi$ and to $\theta$
exponentially vanish at zero frequency as $T\rightarrow 0$, see Fig. 1), it can be shown \cite{mostep3}
that our theory predicts that the Casimir-Lifshitz entropy verifies $S(T=0)=0$.
The same is true for the Casimir-Polder entropy when the plate is a semiconductor.

The ratio of the Casimir-Lifshitz free energies for pure germanium and pure silicon for various conductivity
models is shown in Fig. 2, where the increase of the
energy due to the finite conductivity as compared to the bare permittivity is demonstrated
for large distances. In one case, the theory of drifting
carriers (Eqs.(\ref{refTM},\ref{refTE})) is used to model the interaction of the field with
the plates; in the other, a simple additive term to the bare
permittivity, $4 \pi \sigma_0/\xi$, is employed in the usual Fresnel reflection coefficients \cite{landaulifshitz}.
For the drifting carriers, when the plate separation becomes
much larger that the Debye-H\"uckel screening length, the plates appear as perfect conductors for the TM $n=0$ mode, while in the case of the additive term, the plates appear
as perfect conductors for the TM $n=0$ mode at distances of the order of
$\lambda_T=\hbar c/ k_{\rm B} T$ ($\simeq 7 \mu$m at $T=300$K), independent of the material properties.

Although this effect has yet to be demonstrated for the
Casimir-Lifshitz force, the
drifting carrier treatment provides a way to include a finite conductivity term in the
Casimir-Polder force which has been measured between an atom and a fused silica plate \cite{cornell}.  For fused silica the Debye-H\"uckel screening
length is expected to be extremely large due to the low charge concentration and the quasi-bound character of charges that are contained in a dielectric material. Taking an effective $R_{\rm D}>1$cm is certainly not unreasonable, in which case
the fused silica used in \cite{cornell} can be treated as a perfect
dielectric, as has been assumed in the analysis of this experiment. On the other hand, including the effect of
dc conductivity by adding $4\pi\sigma_0/\xi$ to the permittivity in the usual Fresnel formulas leads to a
increase in the force (by up to nearly a factor of two \cite{mostep2}) at distances of order 10 $\mu$m where the experiment was performed, and this disagrees with the experimental result.

\begin{figure}[t]
\scalebox{0.16}{\includegraphics{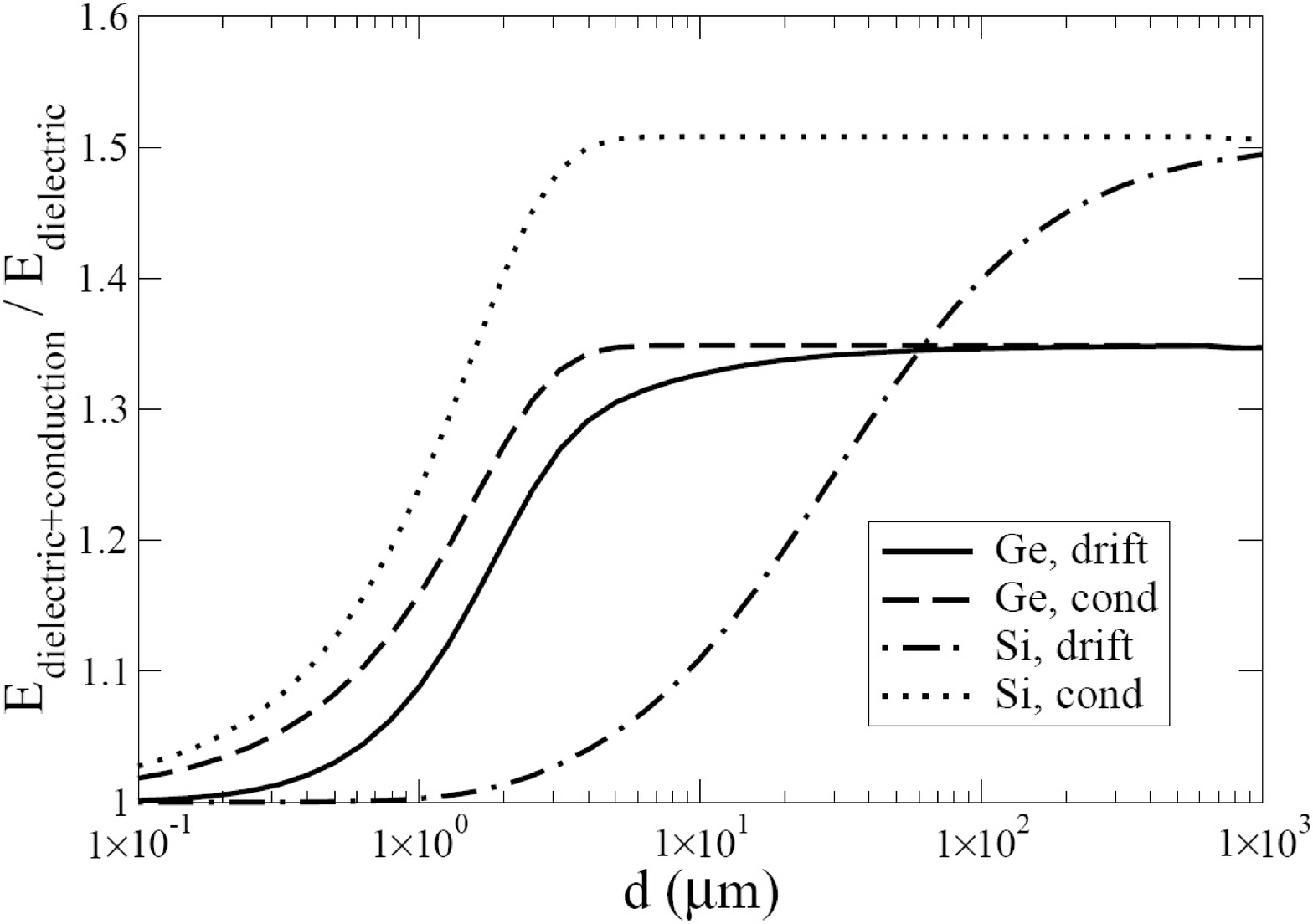}}
\caption{Ratio of Casimir-Lifshitz free energies at $T=300$K for intrinsic semiconductor parallel plates for different conductivity models as follows: Carrier drift with Debye-H\"uckle screening, and a dc conductivity term $4 \pi \sigma_0/\xi$ added to the bare permittivity.
Parameters are as follows:
For Ge, the Debye length is $R_{\rm D}=0.68 \mu$m and the dc conductivity is $\sigma_0=1/(43\ \Omega$ cm);
For Si, $R_{\rm D}=24 \mu$m and $\sigma_0=1/(2.3\times 10^{5}\ \Omega$ cm).
}
\end{figure}



{\it Conclusions.-}
We have shown that treating the finite conductivity of a non-degenerate semiconductor (or insulator) by use of
the classical Boltzmann equation in conjunction with Maxwell's equation
leads to a modification of the Casimir-Lifshitz
force between such materials and provides a way to describe the effects of a small
conductivity. In particular, for small
electric fields such that
$|e E| R_{\rm D} / k_{\rm B} T \ll 1$, as expected for
Casimir and related forces, a standard treatment of adding a term
$4 \pi i\sigma_0/\omega$ to a ``bare" dielectric permittivity is not correct for distances less than the Debye-H\"uckle screening length. This is because the current driven by the electric field,
${\bf J}=\sigma {\bf E}$, is counterbalanced by thermal diffusion, as modelled through the classical Boltzmann equation. Thus, this result represents the dynamic equilibrium between a time-varying field and the charge distribution in the material. However, the finite temperature correction described in \cite{sernelius} and its apparent disagreement with experiment cannot be addressed within the scope of our model which does not apply to metals, where the electron density is sufficiently large that the electron gas is degenerate, so use of the classical Boltzmann equation is not warranted \cite{Ashcroft}.

It is possible to show that the reflection amplitudes derived in this work can be interpreted in terms of ``non-local" dielectric functions (spatial dispersion) \cite{spatialdispersion}.  We have shown that these effects can be derived from readily available material properties, and that
only the quasi-static limit (zero Matsubara frequency TM
mode) is relevant. In the near future we plan to apply
these results to an ongoing measurement of the Casimir-Lifshitz force between pure germanium plates.


We would like to acknowledge correspondence with L.P.  Pitaevskii,
C. Henkel and F. Intravaia,
and discussions with
G.L. Klimchitskaya and V.M. Mostepanenko.

\end{document}